\documentclass[twocolumn,showpacs,nofootinbib,preprintnumbers,amsmath,amssymb,showkeys,superscriptaddress]{revtex4-1}
\usepackage{epsfig}
\usepackage{dsfont}
\usepackage{color}
\usepackage{graphicx}
\usepackage{dcolumn}
\usepackage{bm}
\usepackage{physics}

\newcommand \beq{\begin{equation}}
\newcommand \eeq{\end{equation}}
\newcommand \bea{\begin{align}}
\newcommand \eea{\end{align}}
\newcommand{\nn}{\nonumber \\}

\newcommand\eqn[1]{(\ref{#1})}      
\newcommand\Eqn[1]{Eq.~(\ref{#1})}  
\newcommand\Fig[1]{Fig.~\ref{#1}}  
\newcommand\Sec[1]{Sec.~\ref{#1}}  

\newcommand{\cD}{{\cal D}}

\begin{document}
\allowdisplaybreaks

\title{Backreaction of superhorizon scalar field fluctuations on a de Sitter geometry: \\a renormalisation group perspective}

\author{G. Moreau} 
\author{J. Serreau}
\affiliation{APC, AstroParticule et Cosmologie, Universit\'e Paris Diderot, CNRS/IN2P3, CEA/Irfu, Observatoire de Paris, Sorbonne Paris Cit\'e, 10, rue Alice Domon et L\'eonie Duquet, 75205 Paris Cedex 13, France.}
\date{\today}

\begin{abstract}
We study the backreaction of gravitationally amplified quantum fluctuations of scalar fields on a classical de Sitter geometry. We formulate the problem in the framework of the Wilsonian renormalisation group, which allows us to treat the scalar field fluctuations in a nonperturbative manner and to follow the renormalisation flow of the spacetime curvature as long wavelength, superhorizon fluctuations are progressively integrated out. For light fields in units of the spacetime curvature, these are described by an effective zero-dimensional field theory and can essentially be computed analytically. A nontrivial flow of the spacetime curvature is induced either by a nonminimal coupling to gravity or by self-interactions. The  latter leads to a decrease of the spacetime curvature through loop effects, which, for minimally coupled, massless fields, grow unbounded in the infrared. However, such large loop contributions are eventually screened by the dynamical generation of a nonperturbative, gravitationally induced mass and the renormalisation of the spacetime curvature saturates to a nonzero value. Finally, we show that, in the case of spontaneously broken continuous symmetries, the Goldstone modes do not contribute to the infrared flow of the spacetime curvature, despite being strongly amplified by the gravitational field. 
 \end{abstract}

\maketitle

\section{Introduction}
\label{sec:intro}

Vacuum fluctuations in quantum field theory inevitably contribute to the energy-momentum tensor---in the form of a cosmological constant---and should, thus, be the source of a nontrivial gravitational field. The fact that this widely contradicts actual observations is the statement of the cosmological constant problem \cite{Weinberg:1988cp,Padmanabhan:2002ji,Martin:2012bt}, which actually questions our fundamental understanding of  the interplay between gravity and quantum mechanics. Despite intensive efforts, many groundbreaking proposals and developments over more than half a century, it is fair to say that a fully satisfactory explanation is still missing. 

One line of investigation concerns the possibility that the cosmological constant be dynamically screened by quantum fluctuations (possibly including those of the gravitational field itself) \cite{Polyakov:1982ug,Myhrvold:1983hx,Ford:1984hs,Mottola:1984ar,Antoniadis:1985pj,Mazur:1986et,Antoniadis:1991fa,Tsamis:1992sx}. The scenario is that of semiclassical gravity, with quantum fields self-consistently coupled to a classical (background) gravitational field through Einstein's equations. In the absence of quantum fluctuations, a positive cosmological constant $\Lambda$ sources a maximally symmetric de Sitter geometry, which, in standard cosmological coordinates, corresponds to an exponentially expanding spacetime, with Hubble rate $H\propto\sqrt{\Lambda}$. Now, the dynamics of quantum fields in de Sitter spacetime has been thoroughly investigated, both because the large degree of symmetry permits practical calculations and because of its relevance to inflationary cosmology. Fluctuations of light fields in units of $H$ undergo a dramatic amplification on superhorizon scales, which can be interpreted as tremendous particle production from the gravitational field, similar to Schwinger pair creation from a strong electric field \cite{Mottola:1984ar,Tsamis:2005hd,Krotov:2010ma}. This typically leads to serious infrared issues (infrared and/or secular divergencies) in loop calculations and perturbation theory breaks down on large spacetime scales \cite{Starobinsky:1994bd,Weinberg:2005qc}. Possible exceptions involve symmetries that prevent such infrared contributions, which may be the case of gravitational fluctuations, although this is a subject of debate \cite{Tsamis:1992sx,Senatore:2009cf,Urakawa:2009my,Giddings:2010nc,Miao:2012xc}. In this context, it has been suggested that such large loop contributions may signal an instability of de Sitter spacetime against quantum fluctuations \cite{Polyakov:1982ug,Ford:1984hs,Mottola:1984ar,Antoniadis:1985pj,Mazur:1986et,Antoniadis:1991fa,Tsamis:1992sx,Onemli:2002hr,Tsamis:2005hd,Antoniadis:2006wq,Polyakov:2007mm,Krotov:2010ma,Marolf:2010zp,Hollands:2010pr,Boyanovsky:2011xn,Tsamis:2011ep,Anderson:2013ila,Akhmedov:2012dn,Akhmedov:2013xka,Kaya:2013bga,Akhmedov:2017ooy,Markkanen:2017abw,Anderson:2017hts,Miao:2017vly}. The scenario is that the backreaction of amplified loop contributions leads to an effective decrease of the background field, that is, of the effective cosmological constant. 

To convincingly establish whether this is the case or not is, however, not an easy task. First, the actual calculation of (perturbative) graviton loop contributions is technically involved  \cite{Tsamis:1996qk,Senatore:2009cf,Giddings:2010nc}. Second, even for simple scalar fields, the breakdown of perturbative methods mentioned above requires resummation techniques or genuine nonperturbative approaches. Finally, a proper treatment of a possible instability would require an actual calculation of the dynamics of backreaction away from the maximally symmetric de Sitter geometry, in the veine of  \cite{Boyanovsky:1997cr,Koivisto:2010pj,Markkanen:2013nwa,Glavan:2017jye}, which seriously complicates the calculation of loop contributions. 

In a recent paper \cite{Moreau:2018lmz}, we have proposed a novel approach to the question of backreaction based on nonperturbative renormalisation group (NPRG) techniques in de Sitter spacetime \cite{Kaya:2013bga,Serreau:2013eoa,Guilleux:2015pma,Guilleux:2016oqv,Gonzalez:2016jrn,Prokopec:2017vxx}. This allows us to frame the question in a slightly different way, namely, by studying the build up of quantum fluctuations and of their backreaction as a function not of time but of the renormalisation group (RG) scale. In particular, we can consistently compute the RG trajectories in the subspace of constant field configurations and de Sitter geometries. In that case, we can follow the RG flow of the effective spacetime curvature as the quantum fluctuations of, say, a scalar field are progressively integrated out. Moreover, we can conveniently focus on the role of the superhorizon quantum fluctuations by initialising the RG flow at the horizon scale. 

After presenting the general NPRG formulation of the semiclassical backreaction problem, we apply it to the maximally symmetric case of constant (average) field configurations and de Sitter background geometry in \Sec{sec:framework}. Our goal in this work is to study the backreaction of light scalar fields. In the infrared regime, the dominant contribution to the semiclassical Friedmann equation comes from the effective potential. The infrared RG flow of the latter has been studied in Refs.~\cite{Serreau:2013eoa,Guilleux:2015pma}, where it has been shown to reduce to that of an effective zero-dimensional theory that can be solved essentially analytically. This allows us to analyse the question of backreaction in the present context in a transparent way. 

We discuss the case of Gaussian fields in \Sec{sec:gaussian}, where we show that a nonminimal coupling to gravity alone does induce a nontrivial flow of the spacetime curvature. In \Sec{sec:largeN}, we analyse O($N$)-symmetric theories with quartic self-interaction in the limit $N\to\infty$, where the analytical expressions are particularly simple and transparent, and which qualitatively describe the case of finite $N$ as well. We separately discuss the symmetric and the broken symmetry regimes of the RG flow. In particular, we show that, in the latter case, Goldstone modes, which correspond to flat directions of the effective potential, do not contribute to the flow of the spacetime curvature. This originates from the fact that the correlator of such modes is protected by the underlying symmetry and does not receive any loop correction. Finally, we discuss the general case (finite $N$) in \Sec{sec:N}, again separating the symmetric and broken symmetry regimes. We perform a perturbative analysis, which is valid at the early stages of the flow but eventually breaks down due to unbounded loop contributions. We show how the dynamical generation of a (nonperturbative) mass screens such large fluctuations in the far infrared and stabilises the flow of the spacetime curvature, which eventually saturates at a nonzero value. We summarise our results and present our conclusions in \Sec{sec:concl}.

\section{General framework}
\label{sec:framework}

\subsection{NPRG formulation of the semiclassical problem}

We consider a generic quantum theory of scalar fields ($\hat\varphi$) coupled to gravity ($\hat g_{\mu\nu}$) described by a given microscopic action $S[\hat\varphi,\hat g]$, from which one constructs the effective (quantum) action $\Gamma[\varphi,g]$ as a functional of the average fields $\varphi=\ev{\hat\varphi}$ and $g_{\mu\nu}=\ev{\hat g_{\mu\nu}}$. The latter is the generating functional of one-particle-irreducible vertex functions. It integrates the effects of the quantum fluctuations of both the scalar and the gravitation fields, a complete treatment of which would require a theory of quantum gravity. In the present work, we make two simplifying assumptions. First, we consider energy scales well below the Planck mass $M_P$ so that the effective gravitational coupling is small and one can retain only the lowest order diagrams with gravitational vertices. Second, we assume that loop diagrams involving fluctuations of the gravitational field can be safely neglected, that is, contrarily to diagrams with scalar loops, they are not amplified by infrared effects. Whether this assumption is reliable or not must still be clarified \cite{Tsamis:1992sx,Senatore:2009cf,Urakawa:2009my,Giddings:2010nc,Miao:2012xc}. 

Under the above assumptions, we shall neglect all diagrams with graviton loops, which amounts to treating the gravitational field as a classical (dynamical) geometry, that is, to replacing $\hat g\to\ev{\hat g}=g$, but we aim at taking full account of the scalar loops onto the dynamics, including its backreaction on the background geometry through the field equation $\delta\Gamma/\delta g=0$. This is still a very complicated problem in general. To further simplify matters, we restrict ourselves to constant field configurations $\varphi(x)={\rm const.}$, for which the background gravitational field is described by the maximally symmetric de Sitter geometry (for positive curvature). In this case, it is well-known that the fluctuations of the scalar field undergo a dramatic gravitational amplification on superhorizon scale which yields a nonperturbative infrared dynamics. Our aim is to investigate whether the latter leads to a decay of the spacetime curvature as the scenario discussed above speculates.

Here, we treat the scalar field dynamics using NPRG techniques, following Refs.~\cite{Kaya:2013bga,Serreau:2013eoa,Guilleux:2015pma}, which we adapt to the semiclassical problem at hand. This allows to progressively integrate the infrared fluctuations of the scalar field and to follow the resulting RG flow of the effective curvature of the geometry. Let us first describe the general semiclassical setting. The quantum scalar field theory in the  background metric $g_{\mu\nu}$ is described by the following functional integral
\beq \label{eq:measure}
 e^{-iW_\kappa[J,\,g]}=\int\cD\hat\varphi e^{iS[\hat\varphi,g]+i\Delta S_\kappa[\hat\varphi,g]-iJ\cdot\hat \varphi},
\eeq
where $J\cdot\hat\varphi=\int_xJ(x)\hat\varphi(x)$ and where the quadratic modification of the action
\beq\label{eq:deltaS}
 \Delta S_\kappa[\varphi,g]=\frac{1}{2}\int_{x,y}\!\!R_\kappa(x,y)\varphi(x)\varphi(y)\equiv\frac{1}{2}{\rm Tr}_gR_\kappa[g]\cdot\varphi\varphi
\eeq 
plays the role of an infrared cutoff, which suppresses fluctuations of wavelength larger than $1/\kappa$ (in the sense of the metric $g_{\mu\nu}$) from the path integral.
Here, $\int_x=\int d^D x\sqrt{-g}$  is the invariant measure in $D=d+1$ dimensions and the last equality defines the corresponding functional trace ${\rm Tr}_g$. Note that the scalar regulator function $R_\kappa$ typically depends on the metric.
The regularised effective action $\Gamma_\kappa[\varphi,g]$ is defined through the modified Legendre transformation \cite{Wetterich:1992yh}
\beq
 \Gamma_\kappa[\varphi,g]+\Delta S_\kappa[\varphi,g]+W_\kappa[J,g]=J\cdot\varphi
\eeq
and interpolates between the microscopic action $\Gamma_{\kappa\to\infty}=S$, for $\kappa$ large compared to any other scale in the problem, and the usual effective action $\Gamma_{\kappa\to0}=\Gamma$. One easily shows that
\beq
\partial_\kappa\Big(\Gamma_\kappa[\varphi,g]+\Delta S_\kappa[\varphi,g]\Big)=\Big<\partial_\kappa \Delta S_\kappa[\hat\varphi, g]\Big>,
\eeq
where the average is to be taken with respect to the measure \eqn{eq:measure}, from which one deduces the exact flow equation \cite{Wetterich:1992yh,Delamotte:2007pf} (a dot denotes $\kappa\partial_\kappa$)
\beq\label{eq:Wett}
 \dot\Gamma_\kappa[\varphi,g]=\frac{1}{2}{\rm Tr}_g\dot R_\kappa[g]G_\kappa[\varphi,g],
\eeq
where 
\beq
 G_\kappa[\varphi,g]=i\left(\Gamma_\kappa^{(2)}[\varphi,g]+R_\kappa[g]\right)^{-1}
\eeq
is the exact propagator of the regularised theory, with, defining  the covariant functional derivative as $\delta_c/\delta \varphi(x)=[-g(x)]^{-1/2}\delta/\delta \varphi(x)$,
\beq
 \Gamma_\kappa^{(2)}(x,y)=\frac{\delta_c^2\Gamma_\kappa[\varphi,g]}{\delta\varphi(x)\delta\varphi(y)}.
\eeq
Note that
\beq\label{eq:fieldeom}
  \frac{\delta_c \Gamma_\kappa[\varphi,g]}{\delta \varphi(x)}=J(x)-\int_y R_\kappa(x,y)\varphi(y).
\eeq

So far the settings are just those of a (regularised) field theory in the geometry described by the metric $g_{\mu\nu}$. For our present purposes, the latter is to be determined self-consistently at each scale $\kappa$ from the (exact) extremisation conditions
\beq\label{eq:geom}
   \left.\frac{\delta_c \Gamma_\kappa[\varphi,g]}{\delta \varphi(x)}\right|_{\varphi_\kappa,g_\kappa}\!=0\,,\quad\left.\frac{\delta_c \Gamma_\kappa[\varphi,g]}{\delta g^{\mu\nu}(x)}\right|_{\varphi_\kappa,g_\kappa}\!=0.
\eeq
At the order of approximation considered here for the gravitational fluctuations, this is nothing but the set of (regularised) semiclassical Einstein equations, which encode the backreaction of the scalar field quantum fluctuations onto the average value of the metric field $g_{\mu\nu}$. The second equation in \eqn{eq:geom} writes, equivalently,
\beq
  \left[\left<\frac{\delta_c S[\hat\varphi, g]}{\delta g^{\mu\nu}}\right>_{\!\kappa}+\frac{1}{2} \frac{\delta_c }{\delta g^{\mu\nu}}{\rm Tr}_gR_\kappa[g]G_\kappa[\varphi,g_\kappa]\right]_{g=g_\kappa}\!=0,
\eeq
where the average $\langle\ldots\rangle_{\kappa}$ is evaluated at the extremum $(\varphi_\kappa,g_\kappa)$ and where we stress that the functional derivative in second (regulator) term does not act on $G_\kappa$.

To be more explicit, let us decompose the action in a pure gravitational term and a matter term as
\beq
 S[\varphi,g]=S_{\rm g}[g]+S_{\rm m}[\varphi,g]
\eeq
and define, accordingly,
\begin{align}
 M_P^2G_{\mu\nu}=2\frac{\delta_c S_{\rm g}}{\delta g^{\mu\nu}}\qand T_{\mu\nu}=-2\frac{\delta_c S_{\rm m}}{\delta g^{\mu\nu}}.
\end{align}
 The regularised semiclassical Einstein equations become
\beq
 M_P^2G_{\mu\nu}[g_\kappa]=\Big<T_{\mu\nu}[\hat\varphi,g_\kappa]\Big>_{\!\kappa}+\Delta T^\kappa_{\mu\nu}[\varphi_\kappa,g_\kappa].
\eeq  
The explicit contribution from the regulator reads
\beq\label{eq:regTmunu}
 \Delta T^\kappa_{\mu\nu}(x)=\int_{z,z'}t^\kappa_{\mu\nu}(x;z,z')G_\kappa(z,z'),
\eeq
where we defined
\beq
 t^\kappa_{\mu\nu}(x;z,z')=\qty[g_{\mu\nu}(x)\delta_c(x;z,z')-\frac{\delta_c }{\delta g^{\mu\nu}(x)}]R_\kappa(z,z').
\eeq
with $2\delta_c(x;z,z')=\delta_c(x,z)+\delta_c(x,z')$ and $\delta_c(z,z')=\delta^{(D)}(z-z')/\sqrt{-g(z)}$. As a check, one verifies that a simple mass term, $R_\kappa(z,z')=-m^2\delta_c(z,z')$, yields the expected $\Delta T^\kappa_{\mu\nu}(x)=-m^2g_{\mu\nu}(x)G_\kappa(x,x)/2$.

The general picture is as follows: The progressive integration of the long wavelength scalar field fluctuations through the RG equation \eqn{eq:Wett}, results in an effective renormalisation of the geometry, through the extremization conditions \eqn{eq:geom}. We now specify the above framework to the maximally symmetric case of homogeneous sources, that is, homogeneous field configurations $\varphi$. Assuming that the regulator function can be chosen maximally symmetric as well, one has $\ev{T_{\mu\nu}}_\kappa\propto\Delta T^\kappa_{\mu\nu}\propto g_{\mu\nu}$ and the solution of \Eqn{eq:geom} is a maximally symmetric metric, that is, the de Sitter geometry in the case of a positive curvature, which we consider here.

\subsection{Application to de Sitter space in the infrared limit}

The effective action $\Gamma_\kappa[\varphi,g]$ is defined for arbitrary field/metric configurations. In the following, we consider the hypersurface of maximally symmetric configurations with constant field $\varphi(x)=\varphi$ and de Sitter metric $g_{\mu\nu}(x)=g^{H}_{\mu\nu}(x)$, characterised by a single (Hubble) scale $H$.  More precisely, we shall consider the expanding Poincar\'e patch of the de Sitter geometry. It is always possible---and it proves convenient---to choose the coordinate system where the $H$-dependence of the metric appears as a global rescaling: 
\beq\label{eq:scaling}
 g^{H}_{\mu\nu}(x)=H^{-2}\tilde g_{\mu\nu}(x),
\eeq
where $\tilde g$ is a fiducial de Sitter metric with Hubble parameter $\tilde H=1$. This is, for instance, the case with conformal time $\eta\in\mathds{R}^-$ and comoving spatial coordinates ${\bf X}$, in terms of which the line element reads $ds^2=(-d\eta^2+d{\bf X}^2)/(H\eta)^2$. Using \Eqn{eq:scaling}, we get 
\beq
 H\partial_H\Gamma_\kappa[\varphi,g^H]=2\int_{x} \left.g_H^{\mu\nu}(x)\frac{\delta_c\Gamma_\kappa[\varphi,g]}{\delta g^{\mu\nu}(x)}\right|_{g^H},
\eeq
so that, writing the effective action for constant field as $\Gamma_\kappa[\varphi, g^H]=\int_x V_\kappa(\varphi,H)$, with $V_\kappa$ the effective potential, the second condition in  \Eqn{eq:geom} becomes $\partial_H(H^{-D}V_\kappa)|_{\varphi_\kappa,H_\kappa}=0$. This reduces to the semiclassical Friedmann equation, which defines the effective Hubble parameter $H_\kappa$ renormalized by the fluctuations of the quantum scalar field.

We now consider the flow equation \eqn{eq:Wett}. As discussed in detail in Refs.~\cite{Serreau:2013eoa,Guilleux:2015pma}, the amplified infrared fluctuations result in an effective dimensional reduction: For infrared scales, the RG flow of the effective potential reduces to that of an effective zero-dimensional theory, whose solution at $\kappa=0$ is identical to the late-time equilibrium state of the stochastic approach of Ref.~\cite{Starobinsky:1994bd}. Moreover, it is easy to see, from the Friedmann equation, that the infrared flow of the Hubble parameter is dominated by that of the effective potential. Contributions from kinetic and gradient terms in the energy-momentum tensor are dominated by ultraviolet scales \cite{Boyanovsky:2005sh} and do not contribute to the infrared regime $\kappa\lesssim H_\kappa$. They only affect the initial conditions of the flow, at $\kappa=\kappa_0\sim H_{\kappa_0}$. Finally, it has been shown that the exact effective potential in the infrared limit can be obtained from the lowest order approximation in a derivative expansion of the regularized effective action, known as the local potential approximation (LPA) \cite{Guilleux:2015pma}. 

We consider an O($N$) scalar theory and briefly recall the main features of the LPA and the resulting flow equations. Details can be found in Ref.~\cite{Kaya:2013bga,Serreau:2013eoa,Guilleux:2015pma}. We use the ansatz
\begin{equation}
    \Gamma_{\kappa}[\varphi,g^H] = -\int_x\left\{\frac12 g^{\mu\nu} \partial_\mu\varphi^a \partial_\nu\varphi^a + NU_{ \kappa}(\rho,H) \right\}
    \label{eqn:lpaansatz}
\end{equation}
where $\rho=\varphi^a\varphi^a/(2N)$ and $NU_\kappa=V_\kappa$ is the complete effective potential where we have factored out a $N$ for later purposes. The presence of a kinetic term is dictated by the requirement that $\Gamma_\kappa$ matches the microscopic action in the ultraviolet. However, the running of this term is neglected in the LPA. We work with a regulator function of the form
\beq
 R^{ab}_\kappa(x,x')=-\delta^{ab}\delta(t-t')r_\kappa\qty(|{\bf x}-{\bf x}'|),
 \label{eq:classreg}
\eeq
in terms of the cosmological time $t=-\ln(-\eta)$ and physical coordinates ${\bf x}={\bf X}e^t$. Except for the special case of a pure mass term, $r_\kappa(|{\bf x}|)\propto \delta^{(d)}({\bf x})$---which is a possible infrared regulator but is not enough to regularise the ultraviolet divergences---, the function \eqn{eq:classreg} does not give a fully de Sitter invariant action \eqn{eq:deltaS}. However, the class of regulators \eqn{eq:classreg} is consistent with a large subgroup of de Sitter isometries \cite{Busch:2012ne,Parentani:2012tx,Adamek:2013vw,Guilleux:2016oqv}, which is, in fact, enough in the subspace of constant field configurations.\footnote{The de Sitter breaking effects due to the regulator would only affect the flow of derivative terms, beyond the LPA \cite{Guilleux:2016oqv}.} We choose the function $r_\kappa(|{\bf x}|)=\int\frac{d^dp}{(2\pi)^d}e^{i{\bf p}\cdot{\bf x}} \hat r_\kappa(p)$, with 
\begin{equation}
    \hat r_{ \kappa}(p) =  H^{-D}\left(\kappa^2-p^2H^2\right) \theta\left(\kappa^2-p^2H^2\right),
    \label{eqn:reg}
\end{equation}
which allows for performing the momentum integral in the flow equation analytically and get a simple expression of the beta function of the potential. When combined with the two-point vertex function $\Gamma_\kappa^{(2)}$, the regulator \eqn{eqn:reg} effectively replaces the spatial\footnote{The regulator \eqn{eqn:reg} acts on spatial fluctuations only. An equivalent flow is obtained by coarse graining in the temporal direction in the stochastic approach \cite{Prokopec:2017vxx}, as expected from de Sitter isometries.} gradient term $p^2H^2$ by a mass term $\kappa^2$ for long wavelength modes $p\le \kappa/H$ [all in units of the fiducial scale $\tilde H$; see \Eqn{eq:scaling}]. 

The flow of $U_\kappa$ is obtained by plugging the ansatz \eqn{eqn:lpaansatz} in the exact flow equation \eqn{eq:Wett}, evaluated at constant field configuration.\footnote{It is worth emphasising that the flow equation described here assumes that the quantum field is in the de Sitter invariant Chernikov-Tagirov-Bunch-Davies vacuum state \cite{Chernikov:1968zm,Bunch:1978yq}.} It can be expressed as a sum over a longitudinal and a transverse component 
\begin{equation}
    N \dot{U}_\kappa = \beta(m^2_{l,\kappa},\kappa) + (N-1)\beta(m^2_{t,\kappa},\kappa)
    \label{eq:flowpot}
\end{equation}
where we have omitted the $\varphi$- and $H$-dependences for simplicity and where the longitudinal and transverse curvatures in field space are 
\begin{equation}
    m^2_{l, \kappa}=\partial_\rho U_{ \kappa} +2\rho \partial^2_\rho U_{ \kappa}\qand m^2_{t, \kappa} = \partial_\rho  U_{\kappa}.
    \label{eq:ltmasses}
\end{equation}
In the infrared regime $\kappa\ll H_\kappa$ and for field values where $m_{l/t,\kappa}^2\ll H^2_\kappa$, the beta function takes the simple form
\begin{equation}
    \beta({\cal M}^2,\kappa) =\frac{H^D}{\Omega_{D+1}} \frac{\kappa^2}{\kappa^2 + {\cal M}^2}.
    \label{eq:flowgen}
\end{equation}
Note that, for regions in field space where the potential curvature is not small ${\cal M}^2\gtrsim H_\kappa^2$, the flow is strongly suppressed: $\beta\sim \kappa^{D+1}/{\cal M}$ \cite{Guilleux:2015pma}. As pointed out in Ref.~\cite{Serreau:2013eoa}, the beta function \eqn{eq:flowgen} is similar to that of a zero-dimensional theory. This effective dimensional reduction results from the strong amplification of infrared scalar fluctuations by the gravitational field.

We pause here to note an interesting property of this dimensionally reduced beta function in the case where the field-space curvature at the minimum of the potential vanishes, {\it i.e.}, ${\cal M}^2|_{\rho_\kappa}=0$, which corresponds to a massless, minimally coupled mode of the scalar field. In that case, defining $u_\kappa(H)=\Omega_{D+1}H^{-D}U_\kappa(\rho_\kappa,H)$, we have $\dot u_\kappa(H)=1$. From the definition  $ u_\kappa'(H_\kappa)=0$, we conclude that $\dot H_\kappa=- \dot u_\kappa'(H_\kappa)/u_\kappa''(H_\kappa)=0$. We thus find that, despite being strongly amplified in the infrared (their correlator scales as $ 1/\kappa^2$), the modes associated to a flat direction of the effective potential do not contribute to the renormalisation of the Hubble parameter. Note that this seems to be a robust property of the dimensionally reduced flow \eqn{eq:flowgen}. For instance, the previous argument is insensitive to a possible $H$-dependent redefinition of the square mass term $\kappa^2$ in the regulator function \eqn{eqn:reg}.

As emphasised in Ref.~\cite{Guilleux:2015pma}, we can take another advantage of dimensional reduction in that the solution of the flow equation \eqn{eq:flowpot} is nothing but an effective zero-dimensional field theory whose functional integral representation reduces to a simple integral. Consider the following generating function
\begin{equation}
    e^{N{\cal V}_D\mathcal{W}_\kappa(j,H)} = \int d^N{\hat\varphi} e^{-N{\cal V}_D\left\{U_{\rm in}(\hat\rho,H) + \kappa^2\hat\rho-j\cdot\hat\varphi\right\}}
    \label{eqn:flowsol}
\end{equation}
where ${\cal V}_D=\Omega_{D+1}/H^D$, $\hat\rho=\hat\varphi^2/(2N)$ and $U_{\rm in}$ is to be specified below. It is an easy exercise to check that the regularized effective potential $U_\kappa$, defined as the modified Legendre transform 
\beq
      U_\kappa(\rho,H) + \kappa^2 \rho+\mathcal{W}_\kappa(j,H) = j\cdot\varphi
\eeq
satisfies \Eqn{eq:flowpot} and thus coincide with the effective potential of our initial problem provided one adjusts $U_{\rm in}$ in \Eqn{eqn:flowsol} to match the initial condition at $\kappa=\kappa_0$. In the following, we choose
\beq\label{eq:Uin}
    U_{\rm in}(\rho,H)=a(H)+\mu^2(H)\rho+{\lambda\over2}\rho^2
\eeq
with $a(H)=\alpha-\beta H^2/2+\gamma H^4/4$ and $\mu^2(H)=m^2+\zeta H^2$. The function $a(H)$ stands for the gravitational action evaluated at the de Sitter metric, with the constant and quadratic contributions reflecting the standard Einstein-Hilbert term with a possible cosmological constant, whereas the $H^4$ term describes possible quadratic terms in the curvature tensor, {\it e.g.}, induced by loop effects above the scale $\kappa_0$. We shall see that the latter do not play any role here in $D=4$.  The parameters $\alpha$ and $\beta$ are related to the cosmological constant $\Lambda$ and the Planck mass $M_P$ as $N\alpha=\Lambda M_P^2$ and $N\beta=D(D-1)M_P^2$. The effective square mass function $\mu^2(H)$ includes a possible nonminimal coupling to the Ricci scalar ${\cal R}=D(D-1)H^2$. In terms of the standard normalisation, $m^2+\xi R$, we have $\zeta=D(D-1)\xi$. Again, we have extracted convenient factors of $N$ for later use.

At each scale, the physical values $H_\kappa$ and $\rho_\kappa=\varphi_\kappa^2/(2N)$ are obtained from the extremising conditions
\beq\label{eq:eom}
 \partial_{\varphi_a} U_\kappa=0\qand\partial_H (H^{-D}U_\kappa)=0,
\eeq
which, using the representation \eqn{eqn:flowsol}, are equivalent to the implicit equations $\varphi^a_\kappa = \ev{\hat \varphi^a}_\kappa$ and
\beq\label{eqn:minV}
   \ev{H\partial_H \qty(H^{-D}U_{\rm in})}_{\!\kappa}= DH^{-D}\kappa^2 \qty[\ev{\hat \rho}_{\!\kappa}-\rho_\kappa],
\eeq
where the expectation values are computed with the measure in \eqn{eqn:flowsol} evaluated at $Nj^a=\kappa^2 \varphi_\kappa^a$ and $H=H_\kappa$. Using the explicit expression for $U_{\rm in}$, \Eqn{eqn:minV} rewrites as the following (regulated) semiclassical Friedmann equation, in $D=4$,
\begin{equation}
    \frac{H_\kappa^2}4 = \frac{\alpha + m^2\ev{\hat \rho}_{\!\kappa} + \frac\lambda2  \ev{\hat\rho^2}_{\!\kappa} + \kappa^2[\ev{\hat \rho}_{\!\kappa}-\rho_\kappa]}{\beta - 2\zeta\ev{\hat\rho}_{\!\kappa}},
    \label{eqn:einsteinsc}
\end{equation}
to be compared, {\it e.g.}, with the classical result $H_{\rm cl}^2=4\alpha/\beta$ for a symmetric state with $\rho_\kappa=0$. The last term in the denominator on the right-hand side is the explicit contribution from the regulator, discussed in \Eqn{eq:regTmunu}. Notice, finally, that this is an implicit equation for $H_\kappa$ since the latter enters the expectation values on the right-hand side.

\Eqn{eqn:einsteinsc} can be further simplified by observing that, for a well-behaved function $u$, an integration by part yields
\beq
 \int d^N\hat\varphi \,e^{-u(\hat\varphi)}\hat\varphi^a\partial_{\hat\varphi_a}u(\hat\varphi)=N\int d^N\hat\varphi\, e^{-u(\hat\varphi)}.
\eeq
Applying the latter to \Eqn{eqn:flowsol} and recalling that $j^a=\kappa^2\varphi_\kappa^a/N$, we obtain the identity
\beq
 \ev{\hat\rho\partial_{\hat\rho} U_{\rm in}}_\kappa+\kappa^2\qty[\ev{\hat\rho}_\kappa-\rho_\kappa]=\frac{H_\kappa^D}{2\Omega_{D+1}}.
\eeq
With the choice \eqn{eq:Uin}, we have $\hat\rho\partial_{\hat\rho} U_{\rm in}=\mu^2\hat\rho+\lambda\hat\rho^2$, which allows us to rewrite \Eqn{eqn:einsteinsc} as
\beq
    4\alpha-\beta H_\kappa^2 + \frac{H_\kappa^4}{\Omega}+2\qty(m^2+\kappa^2)\qty[\ev{\hat\rho}_\kappa-\rho_\kappa]+2m^2\rho_\kappa = 0 .
    \label{eqn:minhN}
\eeq
Here $\ev{\hat\rho}_\kappa-\rho_\kappa=G^{aa}_\kappa/(2N)$, with $G_\kappa^{ab}=\ev{\hat\varphi^a\hat\varphi^b}_\kappa-\varphi^a_\kappa\varphi^b_\kappa$ the two-point connected correlator of the theory \eqn{eqn:flowsol}. It is given by, in terms of the effective potential, $G_\kappa^{ab}=(\kappa^2\delta^{ab}+\partial^2_{\varphi^a\varphi^b}U_\kappa)^{-1}$. Decomposing onto longitudinal and transverse components, we have
\beq
 G^{aa}_\kappa=\frac{H_\kappa^D}{\Omega\bar M_{l,\kappa}^2}+(N-1)\frac{H_\kappa^D}{\Omega\bar M_{t,\kappa}^2},
\eeq
with $\bar M_{l/t,\kappa}^2=\kappa^2+m_{l/t,\kappa}^2({\rho_\kappa,H_\kappa})$ the longitudinal/transverse curvatures of the regularised potential at the physical point; see \Eqn{eq:ltmasses}. In conclusion, the equation for $H_\kappa$ is solely governed by the average field $\varphi_\kappa^a$ and the quadratic fluctuations $G^{aa}_\kappa$ around it.

Before presenting some explicit results, let us recall the range of validity of our approach. First, the semiclassical treatment requires that $H_\kappa^2/M_P^2\ll1$, which implies $\alpha/\beta^2\ll1$. Second, the infrared and light field regime of the RG flow require both $\kappa^2\ll H^2_\kappa$ and $m^2_{l/t,\kappa}(\rho,H)\ll H_\kappa^2$. As already emphasized, the flow is strongly suppressed for values of the field where this last condition is not fulfilled.

\section{Gaussian theory}
\label{sec:gaussian}


We first consider the case of a Gaussian theory, {\it i.e.}, with $\lambda=0$, where the effective potential is one-loop exact.
Defining $\mu_\kappa^2(H)\equiv \mu^2(H) + \kappa^2$, we get
\beq
     U_\kappa = a + \mu^2\rho + \frac{H^D}{2\Omega_{D+1}}\ln\frac{{\Omega_{D+1}\mu_\kappa^2}}{2\pi H^D},
    \label{eq:UGauss}
\eeq
where we omitted the implicit $H$-dependences of the functions $a$, $\mu^2$, and $\mu_\kappa^2$ for simplicity.\footnote{As a side remark, we mention that the apparent mismatch in dimension under the logarithm simply reflects the necessity to properly take care of the dimension of the field variable $\hat\varphi$ when performing the Gaussian integral to define a dimensionless partition function . \Eqn{eq:UGauss} must be understood with a factor $\mu_0^{D-2}$ under the logarithm, where $\mu_0$ is an arbitrary mass scale.} The correction to the classical potential $U_{\rm in}=a+\mu^2\rho$ arises from the integration over Gaussian quantum fluctuations and are controlled by the quantity ${\cal V}_D^{-1}=H^D/\Omega_{D+1}$, as clear from \Eqn{eqn:flowsol}.

\begin{figure}[t]
    \centering
    \includegraphics[width=.475\textwidth]{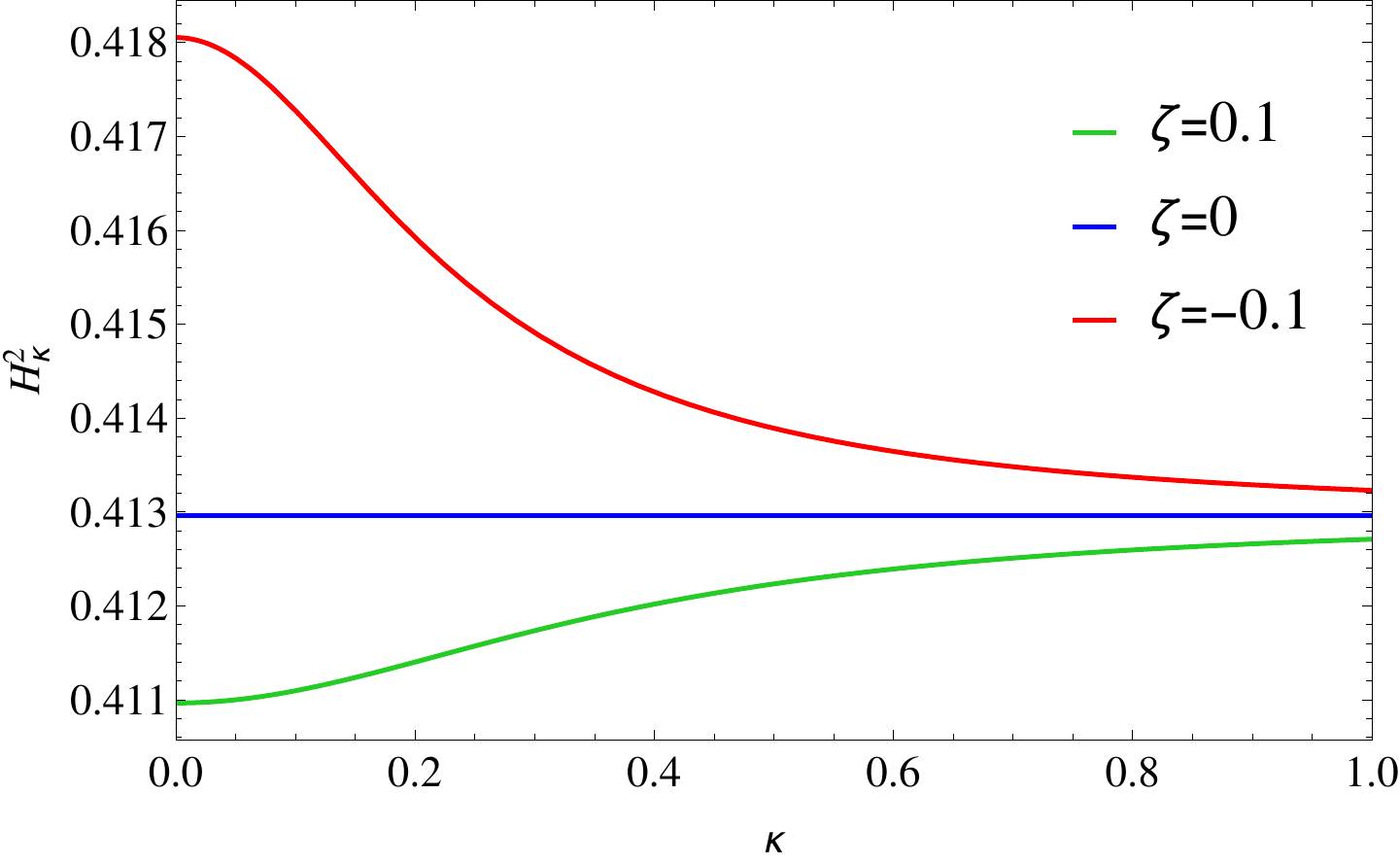}
    \caption{Flow of $H_\kappa^2$ in the Gaussian case ($\lambda=0$) for $m^2=0.1$ and different (positive and negative) values of the nonminimal coupling $\zeta$. The gravitational parameters are $\alpha=0.1$ and $\beta=1$.}
    \label{fig:Gaussian}
\end{figure}

The system \eqref{eq:eom} gives $\rho_\kappa=0$ and, in $D=4$, 
\begin{equation}
    4\alpha- \beta H_\kappa^2 + \frac{H_\kappa^4}{\Omega}\qty( 1+\frac{m^2+\kappa^2}{\bar\mu_\kappa^2}) = 0,
    \label{eqn:hgaussian}
\end{equation}
with $\Omega\equiv\Omega_5=8\pi^2/3$ and $\bar\mu_\kappa^2=\mu_\kappa^2(H_\kappa)=m^2+\kappa^2+\zeta H_\kappa^2$. We thus check that this direct calculation agrees with \Eqn{eqn:minhN}. The last two terms on the left-hand side arise from the logarithmic (loop) correction in \Eqn{eq:UGauss}. When $\zeta=0$, \Eqn{eqn:hgaussian} does not depend explicitly on $\kappa$ and there is no flow, as expected. That is because the only effect of the regulator in that case is to renormalise the $H^4$ term in the function $a(H)$, which plays no role in the semiclassical Friedmann equation. In that case, we have
\begin{equation}
    4\alpha\Omega - \beta\Omega H^2_{\kappa} + 2H^4_{\kappa} = 0
    \label{eqn:masslessIRinitial}
\end{equation}
and the regime of validity of our approach selects the solution with $H^2/\beta\ll1$, that is,
\begin{equation}
    H^2_{\kappa} = \frac{\beta\Omega}4\qty( 1 - \sqrt{1-\frac{32\alpha}{\beta^2\Omega}} )\approx H^2_{\rm cl}+\frac{2H^4_{\rm cl}}{\beta\Omega}.
    \label{eq:UV}
\end{equation}
Here, $H_{\rm cl}^2=4\alpha/\beta=\Lambda/3$ is the classical solution and the second term on the right-hand side is the first quantum correction at the (ultraviolet) scale $\kappa_0$.
Instead, a nonminimal coupling $\zeta\neq0$ induces a nontrivial flow already in the Gaussian theory, as illustrated in \Fig{fig:Gaussian}. The sign of the flow is controlled by that of $\zeta$ and we see that $\zeta<0$ increases the spacetime curvature.\footnote{Note, though, that the Gaussian theory is only well defined if $m^2+\zeta H^2_\kappa>0\,\,\,\forall \kappa$.}

We also note that, for $\alpha=0$, the solution $H_\kappa=0$, corresponding to Minkowski space, is a fixed point of the RG flow \eqn{eqn:hgaussian}. Although appealing, this has to be taken with a grain of salt because, strictly speaking, the above flow equations are only valid for $H_\kappa\neq0$ since, in particular, they rely on approximations such as $\mu^2\ll H^2$, etc. Still, it is an important property, which guarantees, for instance, that $H_\kappa^2$ cannot change sign along the flow. We shall see below that this remains true for interacting theories.

\section{Interacting theory: $N\to\infty$}
\label{sec:largeN}

In the case of interacting theories, $\lambda\neq0$, the equations \eqn{eq:eom} can easily be solved numerically. One can, however, gain physical insight by considering the limit $N\to\infty$, where the running effective potential can be computed analytically from \Eqn{eqn:flowsol} using the saddle point method \cite{Guilleux:2015pma}. We get, after some calculations,
\begin{equation}
        U_\kappa+ \kappa^2 \rho = a + \frac{ M_\kappa^4 - \mu_\kappa^4}{2\lambda} + \frac{H^D}{2\Omega_{D+1}}\ln\frac{\Omega_{D+1} M^2_\kappa}{2\pi eH^D} ,
    \label{eqn:potentiallargeN}
\end{equation}
where, again the implicit $\rho$ and $H$ dependences have been omitted for simplicity. Here, $\mu_\kappa^2\equiv\mu_\kappa^2(H)$ has been defined in \Eqn{eq:UGauss} and \begin{equation}
     M^2_\kappa = \frac{\mu_\kappa^2+ \lambda\rho}2 + \sqrt{\qty(\frac{\mu_\kappa^2 + \lambda\rho}2)^2 + \frac{\lambda H^D}{2\Omega_{D+1}}}
    \label{eqn:zbar}
\end{equation}
is the transverse curvature of the regularized potential, $M^2_\kappa=\partial_\rho U_\kappa+\kappa^2$. 
One recovers the Gaussian result \eqn{eq:UGauss} in the limit \smash{$\lambda\to0$}. Also, as before, formally sending the loop parameter $H^D/\Omega_{D+1}\to0$, one recovers the classical result $U_\kappa=U_{\rm in}$.

The second extremization conditions \eqn{eq:eom} yield, for the running Hubble parameter,
\begin{equation}
    4\alpha-\beta H_\kappa^2 + \frac{2}{\lambda}\qty(\bar M^2_\kappa - \bar\mu_\kappa^2)\qty(\bar M^2_\kappa + m^2 + \kappa^2)  = 4\kappa^2\rho_\kappa
    \label{eqn:minh0}
\end{equation}
with $\bar M^2_\kappa=M^2_\kappa(\rho_\kappa,H_\kappa)$. Equivalently, one can deduce the same equation by explicitely evaluating the various averages entering \Eqn{eqn:einsteinsc} in the large-$N$ limit. Finally, we can simplify this equation by using the equation of motion for $\rho_\kappa$, as we have done to arrive at \Eqn{eqn:minhN}. One has to distinguish two cases for the solution of the field equation $\partial_{\varphi_a}U_\kappa=\varphi_a\partial_\rho U_\kappa/N=0$: either $\varphi_\kappa^a$, and thus $\rho_\kappa$, vanish (symmetric regime), or it is determined by $\partial_\rho U_\kappa|_{\rho_\kappa}=0$ in the broken symmetry regime. 
It is easy to check that \Eqn{eqn:minh0} rewrites, in both cases, as
\beq
    4\alpha-\beta H_\kappa^2 + \frac{H_\kappa^4}{\Omega}\qty(1+\frac{m^2+\kappa^2}{\bar M^2_\kappa})+2m^2\rho_\kappa=0
    \label{eqn:minh}
\eeq 
which reproduces \Eqn{eqn:minhN} in the limit $N\to\infty$, as expected. We shall discuss the symmetric and broken symmetry regimes separately.

\subsection{Symmetric regime}

\begin{figure}[t]
    \centering
    \includegraphics[width=.48\textwidth]{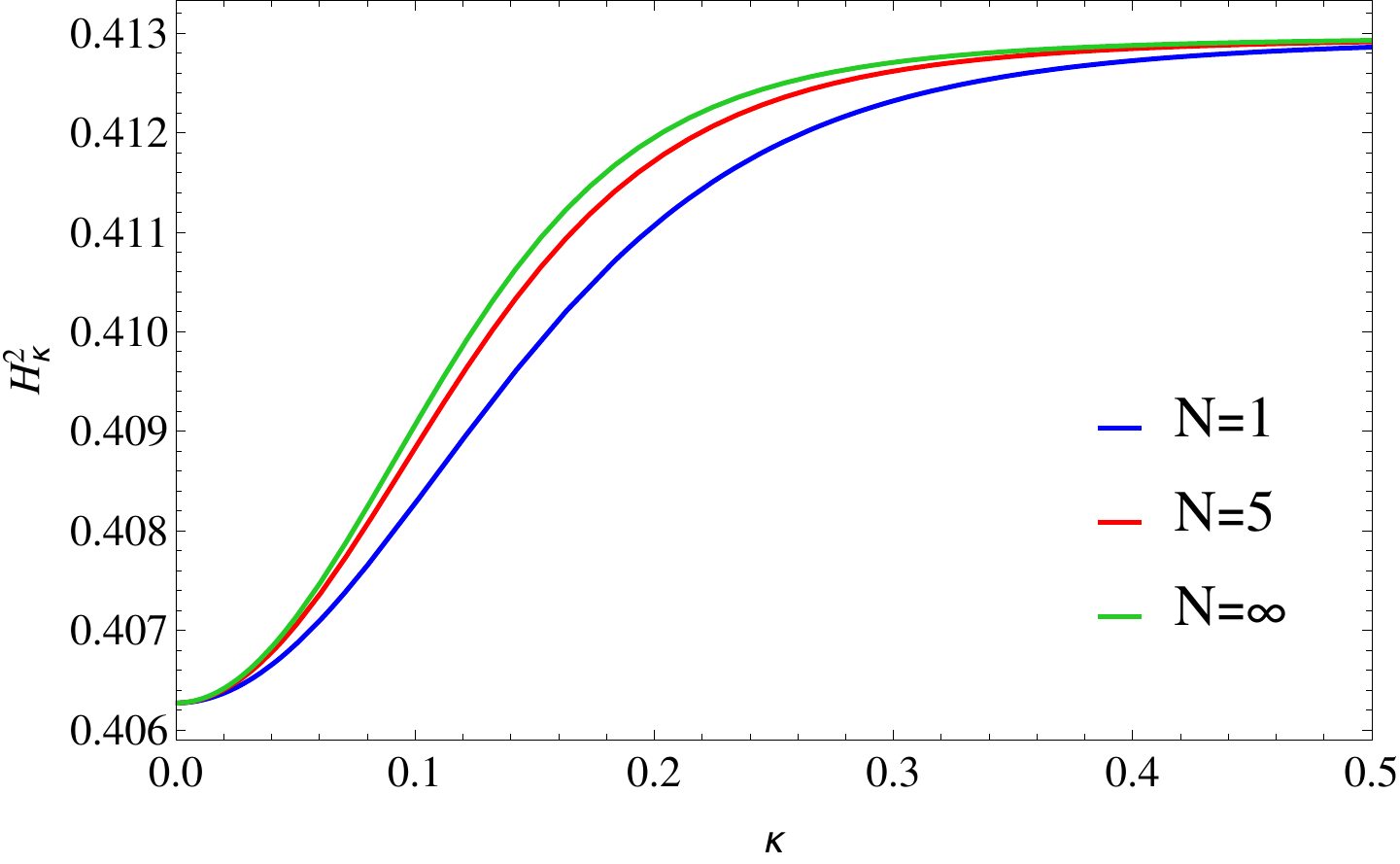}
    \caption{Flow of $H_\kappa^2$ for the interacting massless, minimally coupled theory ($m^2=\zeta=0$), with $\lambda=0.1$, for various values of $N$. The gravitational parameters are $\alpha=0.1$ and $\beta=1$.}
    \label{fig:massless}
\end{figure}

Here, $\rho_\kappa=0$ and the relevant equation becomes
\begin{equation}
    4\alpha-\beta H_\kappa^2 + \frac{H_\kappa^4}{\Omega}\qty(1+\frac{m^2+\kappa^2}{\bar M^2_\kappa}) = 0 
    \label{eq:Hsymeq}
\end{equation}
with 
\begin{equation}
    \bar M^2_\kappa = \frac{\bar \mu_\kappa^2}2 + \sqrt{\qty(\frac{\bar \mu_\kappa^2}2)^2 + \frac{\lambda H_\kappa^4}{2\Omega}},
    \label{eq:barmass}
\end{equation}
which makes the analysis of the flow of $H_\kappa$ particularly simple. One illustration of interest is the massless, minimally coupled field, $m^2=\zeta=0$, whose dynamics is genuinely nonperturbative even at weak coupling. The flow of $H_\kappa^2$ for this case is shown in \Fig{fig:massless}. 

For large enough $\kappa$, we have essentially a Gaussian theory with a large mass $M^2_\kappa \approx \kappa^2$ and \Eqn{eq:Hsymeq} reduces to \Eqn{eqn:masslessIRinitial} with the solution \eqn{eq:UV}.
For intermediate scales $\kappa^2\sim \sqrt{\lambda/\Omega}H_{\kappa_0}^2$, the selfinteraction term in \Eqn{eq:barmass} induces a negative renormalisation of $H_\kappa$ due to infrared enhanced loop effects, as discussed below in \Sec{sec:N}. These are eventually screened by the dynamical generation of a nonvanishing mass $\bar M_{\kappa=0}^2= \sqrt{\lambda/(2\Omega)}H_{\kappa=0}^2$ and, for scales $\kappa^2\ll \bar M_{\kappa=0}^2$, the flow freezes. \Eqn{eq:Hsymeq} becomes
\begin{equation}
    4\alpha\Omega - \beta\Omega H_{\kappa=0}^2 +H^4_{\kappa=0} = 0,
    \label{eqn:masslessIR}
\end{equation}
with the relevant solution 
\beq
  H^2_{\kappa=0}= \frac{\beta\Omega}2\qty( 1 - \sqrt{1-\frac{16\alpha}{\beta^2\Omega}} )\approx H^2_{\rm cl}+\frac{H^4_{\rm cl}}{\beta\Omega}.
    \label{eq:IR}
\eeq
Interestingly, the infrared value $H_{\kappa=0}^2$ is independent of the coupling, which is clear from \Eqn{eq:Hsymeq} and the fact that $\bar M_{\kappa=0}^2$ is nonzero. Comparing the two asymptotic values \eqn{eq:UV} and \eqn{eq:IR}, we conclude that the renormalisation of $H^2_\kappa$ due to infrared fluctuations is controlled by $H^2_{\rm  cl}/(\beta\Omega)\ll1$.

\begin{figure}[t]
    \centering
    \includegraphics[width=.485\textwidth]{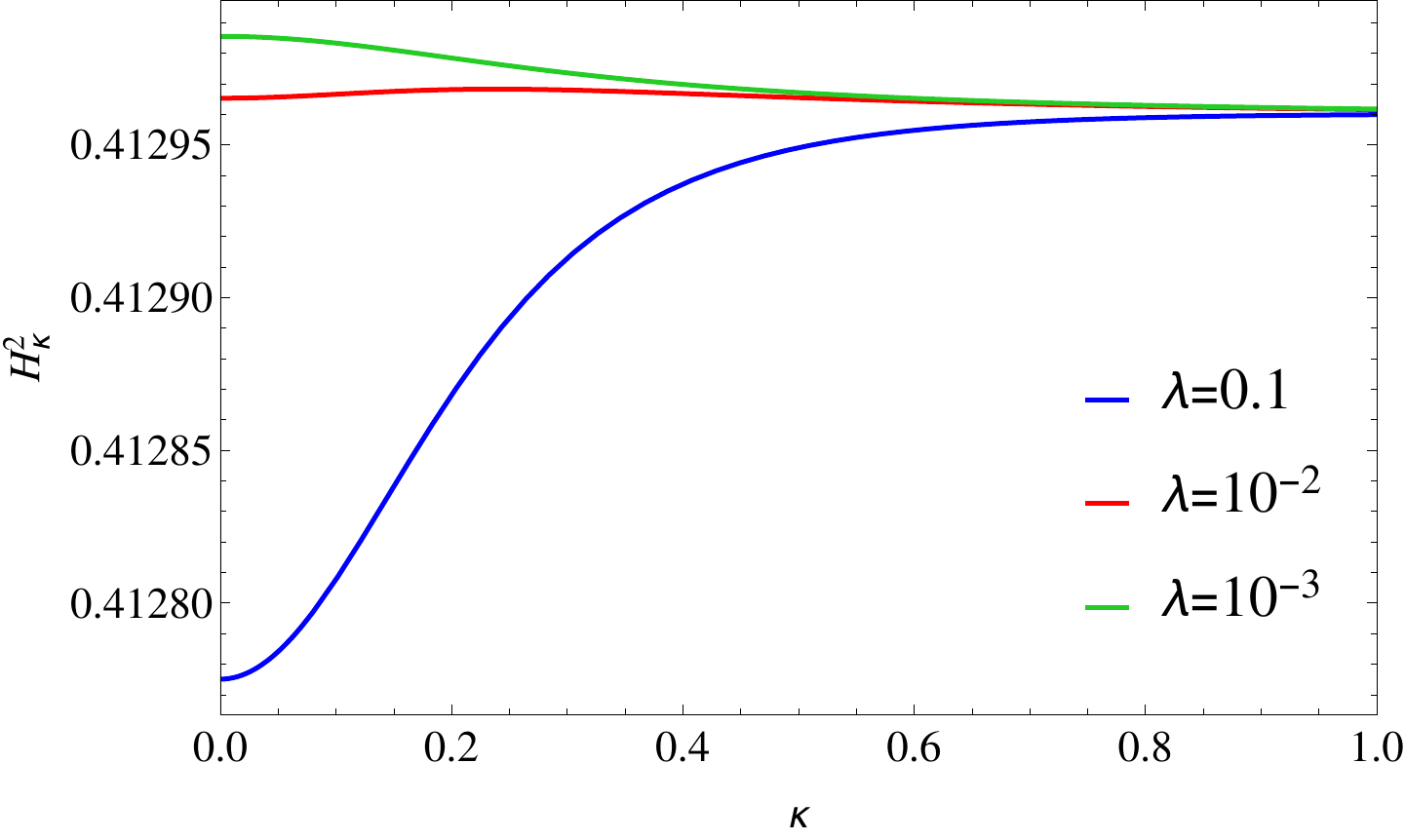}
    \caption{Interplay between the positive renormalisation of $H_\kappa^2$  (as $\kappa\to0$) induced by a negative value of $\zeta$ and the negative renormalisation induced by $\lambda>0$ for a flow in the symmetric regime, $\rho_\kappa=0$. The parameters are $N=\infty$, $m^2=0.1$, and $\zeta=-10^{-3}$. The gravitational parameters are $\alpha=0.1$ and $\beta=1$.}
    \label{fig:interplay}
\end{figure}
The analysis of the general case, $m^2\neq0$ or $\zeta\neq0$, goes along similar lines and the overall picture is the same as above for $\zeta\ge0$. For negative values, there is an interplay between the effects of positive renormalisation induced by $\zeta<0$ and the negative one induced by $\lambda>0$, as illustrated in \Fig{fig:interplay}.

\subsection{Broken symmetry regime}

We now consider the case where the flow is initialised in a state of broken symmetry at the scale $\kappa_0$.\footnote{At the classical level, this requires $m^2+\zeta H_{\kappa_0}^2<0$. We shall see below that, with quantum corrections, this condition becomes $m^2+\zeta H_{\kappa_0}^2<-\lambda H_{\kappa_0}^4/(2\Omega\kappa_0^2)$.} In this regime, the value of $\rho_\kappa>0$ is determined from $\partial_\rho U_\kappa=0$. In the $N\to\infty$ limit, the flow of $\rho_\kappa$ and $H_\kappa$ is fully driven by the massless Goldstone modes and is particularly simple. Indeed, we have $\bar M^2_\kappa = \kappa^2$ and \Eqn{eqn:minh} reduces to
\beq
   4\alpha'\Omega - \beta'\Omega H_\kappa^2 + 2H_\kappa^4 = 0,
    \label{eqn:brokenIRinitial}
\eeq
where $\alpha'=\alpha-(m^2)^2/(2\lambda)$ and $\beta'=\beta+2\zeta m^2/\lambda$. The coefficients of this equation being independent of $\kappa$, we conclude that $H_\kappa=H_{\kappa_0}$ has no flow in this regime. This is an explicit example of the discussion below \Eqn{eq:flowgen}: Goldstone modes do not contribute to the flow of $H_\kappa$. It is instructive to see explicitly how this happens in the present case. The value of the potential at its minimum runs as 
\beq
 U_\kappa(\rho_\kappa,H)=a(H)-\frac{\qty[\mu^2(H)]^2}{2\lambda}+\frac{H^D}{2\Omega_{D+1}}\ln\frac{\Omega_{D+1} \kappa^2}{2\pi H^D}.
\eeq
The first two terms on the right-hand side give the classical value whereas the quantum correction is all contained in the last term, which merely corresponds to the (one-loop) contribution from Gaussian fluctuations of mass $\kappa^2$; see \Eqn{eq:UGauss}. So, despite being strongly amplified, the massless Goldstone modes only yield a slight logarithmic running of the potential at its minimum in the form of a term $\propto H^D$, that is, in $D=4$, a renormalisation of the parameter $\gamma$ in $a(H)$, which does not contribute to the equation for $H_\kappa$.

\begin{figure}[t]
    \includegraphics[width=.48\textwidth]{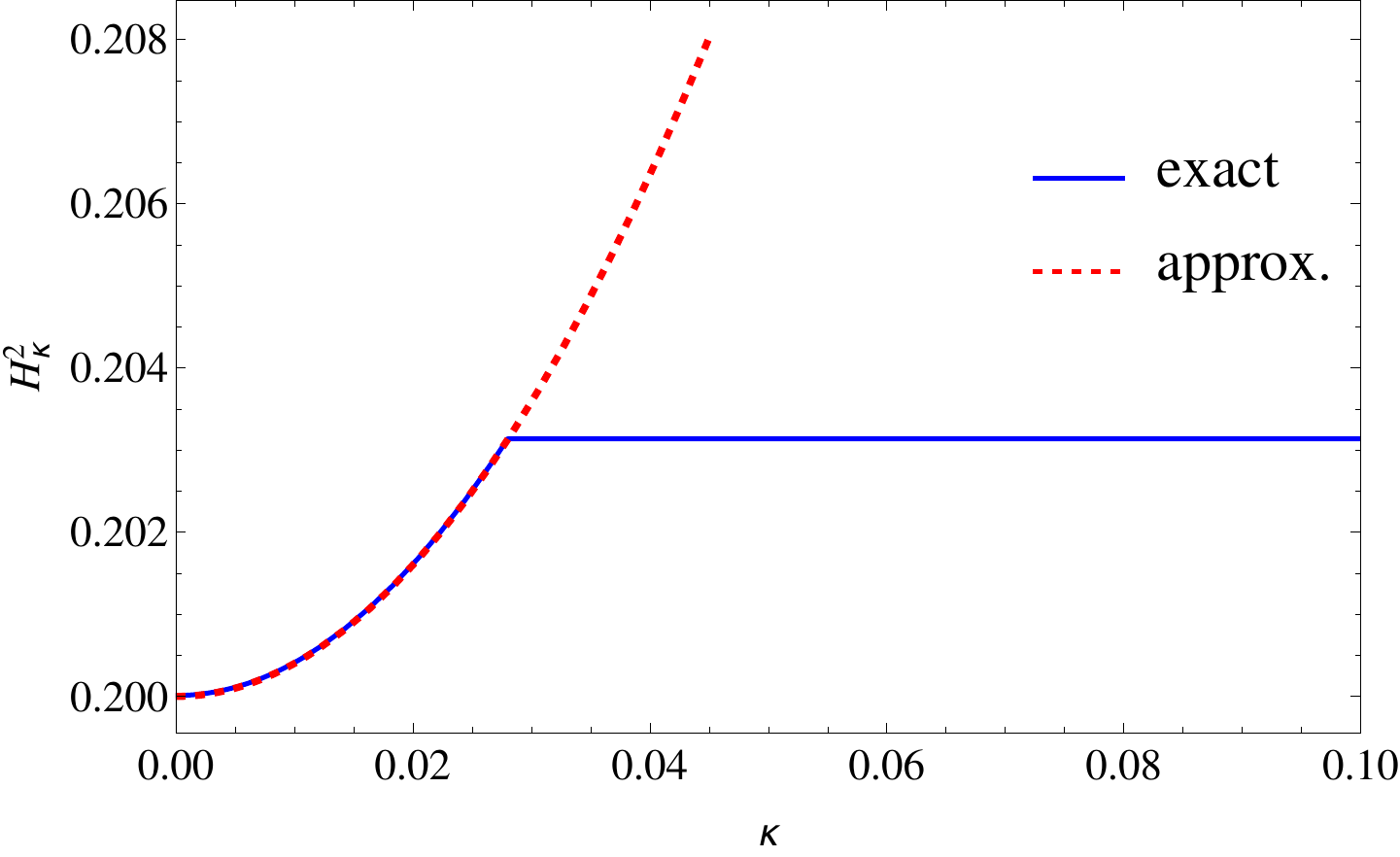}\\
   \vspace{.5cm}
   \hspace*{.32cm}\includegraphics[width=.458\textwidth]{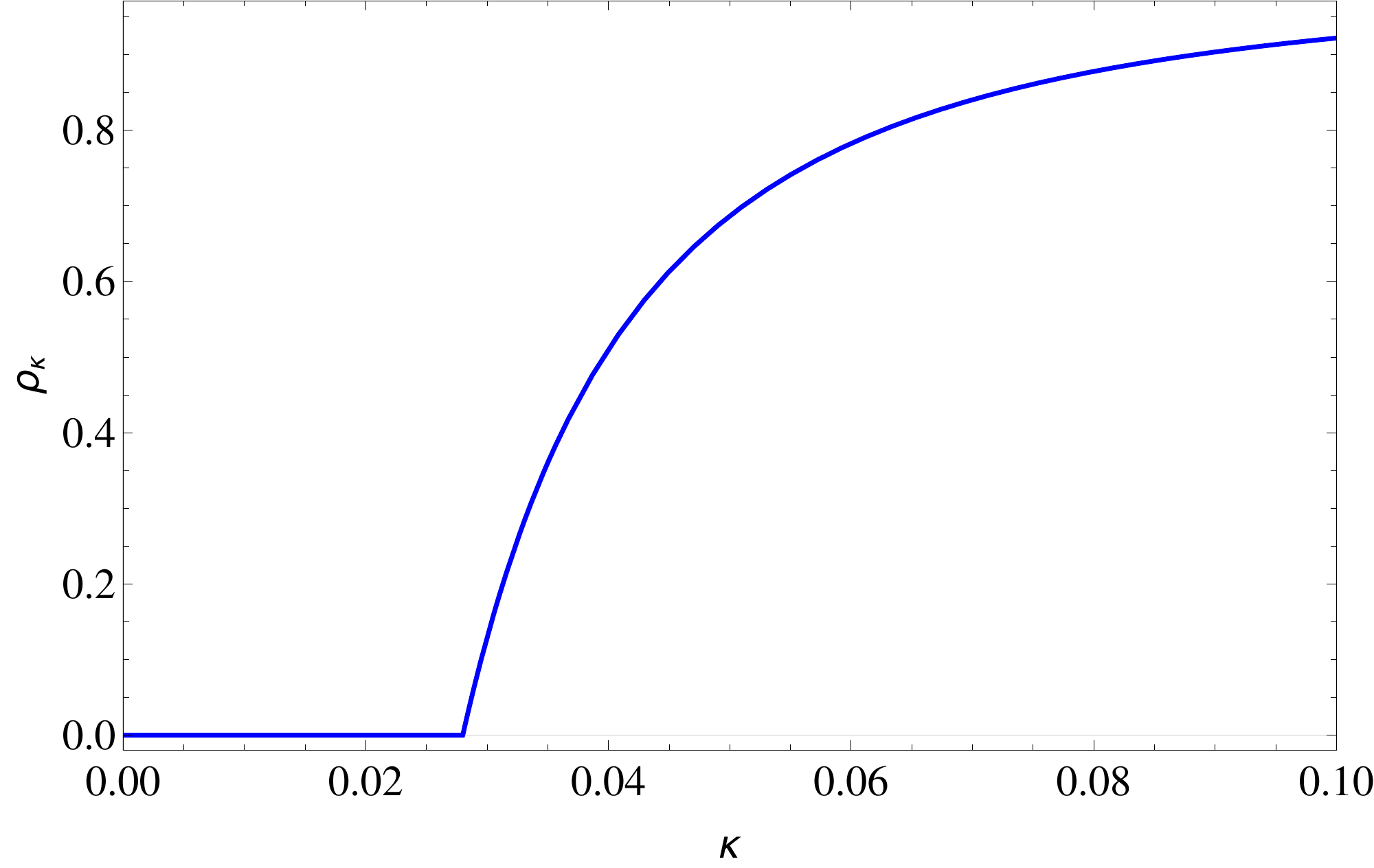}
    \caption{Flows of $H_\kappa^2$ (top) and of $\rho_\kappa$ (bottom) for initial conditions in the broken symmetry regime for $N=\infty$. The parameters are $m^2=-0.1$, $\zeta=0$, and $\lambda=0.1$. The gravitational parameters are $\alpha=0.1$ and $\beta=1$. In the broken symmetry regime, where $\rho_\kappa>0$, the flow is governed by the Goldstone modes which, as explained in the text, do not yield any flow of $H_\kappa$. The symmetry gets restored at a finite RG scale and the remaining flow is that of the symmetric regime. For the parameters used here, the flow of $H_\kappa$ in the symmetric regime is well-described by the approximation \eqn{eq:approxSB}, as shown by the dashed curve.}
    \label{fig:rolegoldstone}
\end{figure}

 \Eqn{eqn:brokenIRinitial} is just the same as \Eqn{eqn:masslessIRinitial} in the symmetric regime with $(\alpha,\beta)\to(\alpha',\beta')$. Here, the space of solutions within the range of applicability of our approach, $0<H_\kappa^2\ll\beta$ is larger than before because $\alpha'$ and $\beta'$ can take different values, either positive or negative. We restrict to the solution continuously related to the one in the symmetric regime in the classical theory, which only exists for $\alpha'>0$ and $\beta'>0$ and is given by \Eqn{eq:UV} with the appropriate replacements. In particular the classical solution is now $H_{\rm cl}^{\prime2}=4\alpha'/\beta'$. For generic choices of parameters, the approximate solution in \Eqn{eq:UV} is valid, although it is possible to fine-tune the parameters so that the quantum corrections be large, {\it i.e.}, $32 \alpha'/(\beta^{\prime2}\Omega)\sim1$, while still having $0<H^2_{\kappa_0}\ll\beta$ (see below).
 
The flow of $\rho_\kappa$ in the broken symmetry regime is also easily deduced from $\bar M_\kappa^2=\kappa^2$. Using that $H_\kappa=H_{\kappa_0}$ in this regime, we get \cite{Guilleux:2015pma}
\beq
\rho_\kappa =  - \frac{m^2+\zeta H_{\kappa_0}^2}\lambda - \frac{H_{\kappa_0}^4}{2\Omega\kappa^2}.
\eeq
We now see the precise condition on the parameters for the flow to start in the broken symmetry regime, given by $\rho_{\kappa_0}>0$. As pointed out in \cite{Serreau:2013eoa,Guilleux:2015pma}, one important consequences of the dimensionally reduced flow is that the symmetry eventually gets restored at the finite scale $\kappa_*$, given by $\rho_{\kappa_*}=0$, that is,
\beq
 \kappa_*^2=\frac{\lambda H_{\kappa_0}^4}{2\Omega|m^2+\zeta H_{\kappa_0}^2|}.
\eeq
The length scale $1/\kappa_*$ can be viewed as the spatial size of domains of broken symmetry. For larger length scale, these domains add incoherently and the symmetry is effectively restored. The subsequent flow, for $\kappa\le\kappa^*$, is that of the symmetric regime discussed in the previous section, Eqs.~\eqn{eq:Hsymeq} and \eqn{eq:barmass}. These simplify even further for a flow initialised sufficiently deep in the broken symmetry regime. For $\lambda H_{\kappa_0}^4\ll|m^2+\zeta H_{\kappa_0}^2|^2$, the running square mass reads
\beq
 \bar M_{\kappa}^2=\frac{\lambda H_{\kappa}^4}{2\Omega|\bar\mu_{\kappa}^2|}\qty[1-\frac{\lambda_{{\rm eff},\kappa}}{2}+{\cal O}\qty(\lambda_{{\rm eff},\kappa}^2)]
\eeq
where we have defined $\lambda_{{\rm eff},\kappa}\equiv\lambda H_{\kappa}^4/\qty(\Omega|\bar\mu_{\kappa}^2|^2)$, and \Eqn{eq:Hsymeq} becomes 
 \begin{equation}
    4\alpha_\kappa'- \beta_\kappa' H_\kappa^2 + \frac{H_\kappa^4}{\Omega}\qty( 1-\frac{m^2+\kappa^2}{\bar\mu_\kappa^2}) = 0,
    \label{eqn:hSB}
\end{equation} 
where $\alpha_\kappa'=\alpha-(m^2+\kappa^2)^2/(2\lambda)$ and $\beta'_\kappa=\beta+2\zeta(m^2+\kappa^2)/\lambda$. This is similar to \Eqn{eqn:hgaussian} for the Gaussian case with $(\alpha,\beta)\to(\alpha'_\kappa,\beta'_\kappa)$ and with a change of sign in the parenthesis. This rewrites in the simpler form
\beq
 4\alpha_\kappa'- \beta_\kappa' H_\kappa^2 + \frac{\zeta H_\kappa^6}{\Omega\bar\mu_\kappa^2} = 0.
\eeq
 In particular, in cases where $|\zeta H_\kappa^2|\ll|m^2|$, we have  
\beq\label{eq:approxSB}
 H_\kappa^2\approx\frac{4\alpha_\kappa'}{\beta'_\kappa}.
\eeq 
Note that the late time result in that case is given by the classical solution $H_{\kappa=0}^2=H_{\rm cl}^{\prime2}$.
  
\begin{figure}[t]
    \centering
    \includegraphics[width=.48\textwidth]{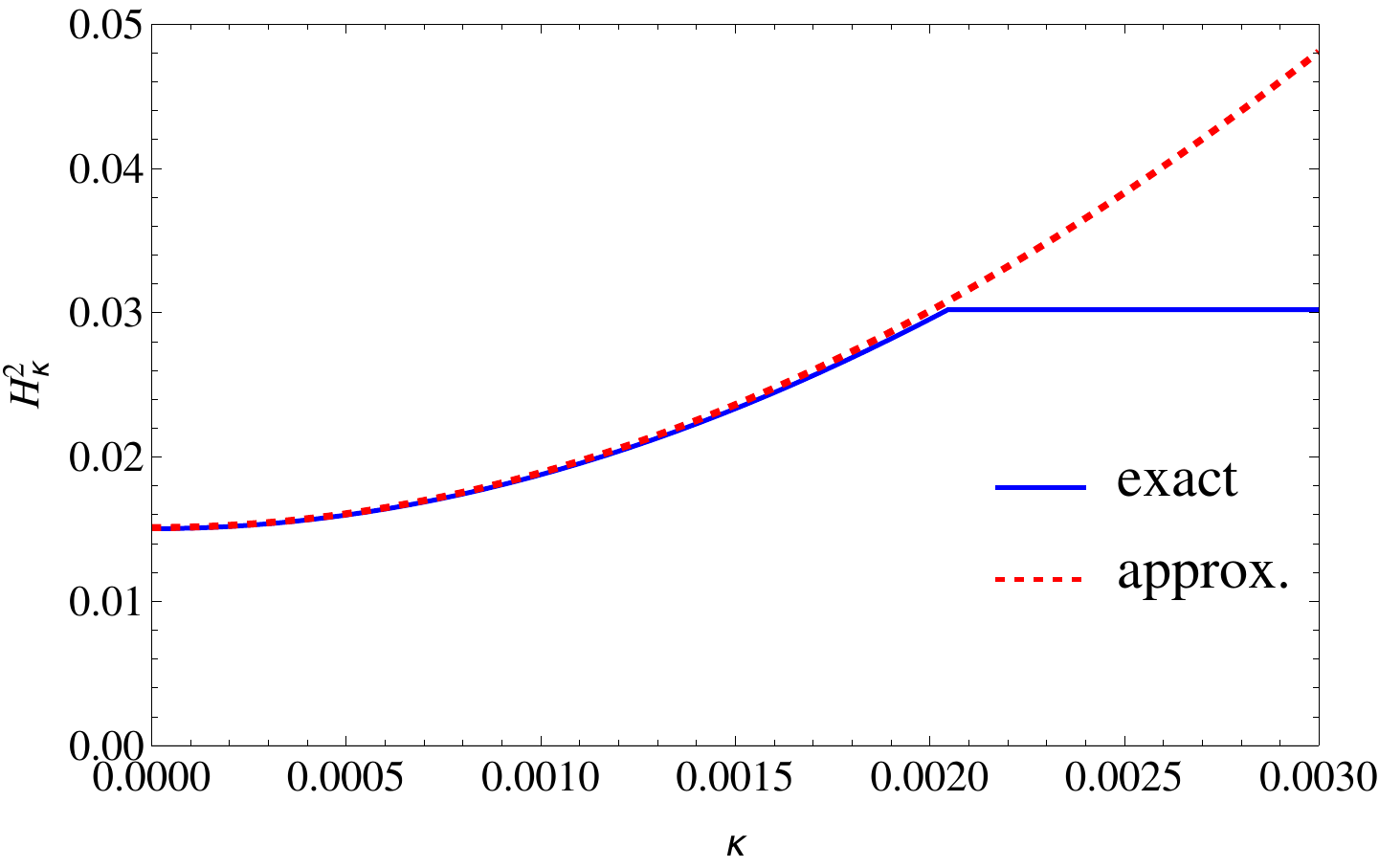}
    \caption{Flow of $H_\kappa^2$ for initial conditions in the broken symmetry regime for $N=\infty$. Here, the parameters are fine-tuned such as to maximise the infrared renormalization effects (see footnote~\ref{ft:foot}): $m^2=-0.0447175$, $\zeta=0.1113$, and $\lambda=0.01$. The gravitational parameters are $\alpha=0.1$ and $\beta=1$.}
    \label{fig:finetune}
\end{figure}

  For generic parameters which satisfy all the constraints arising from initialising the flow in the broken symmetry regime and from the range of validity of our approach, the quantum correction to the classical term $H_{\rm cl}^{\prime2}$ is small and the relative change in $H_{\kappa}^2$ is controlled by $H_{\rm cl}^{\prime2}/(\beta'\Omega)\ll1$. As mentioned before, one can fine-tune the parameters so that the quantum corrections in Eqs.~\eqn{eq:UV} and \eqn{eq:IR} be large, still respecting the range of validity of our approximations. In that case, the relative change in $H_\kappa^2$ after symmetry restoration can be significant but it never exceeds a factor two.\footnote{\label{ft:foot}The existence of a solution continuously related to the classical solution requires $\alpha'>0$, $\beta'>0$, and $\Omega\beta'^2>32\alpha'$. Large deviations from the classical solution require the discriminant to be small $0<1-32\alpha'/(\Omega\beta'^2)\ll1$, in which case $H_{\kappa_0}^2\approx\beta'\Omega/4$. The validity of the semiclassical approximation then imposes $\alpha'\sim\beta'^2\ll\beta^2$. Altogether this means $m^2/\sqrt{\lambda}\gtrsim-\sqrt{2\alpha}$ and $\zeta/\sqrt{\lambda}\lesssim\beta/(2\sqrt{2\alpha})$. For small enough coupling $\lambda$, we can always ensure that the mass term $|m^2+\zeta H_{\kappa_0}^2|\ll H_{\kappa_0}^2$. Further adjusting the parameters so that \Eqn{eq:approxSB} is valid, we obtain $H_{\kappa=0}^2\approx H_{\kappa_0}^2/2$.} This is illustrated in \Fig{fig:finetune}.

\section{Finite $N$}
\label{sec:N}

Similar results can be obtained for any finite value of $N$, where the two equations \eqn{eq:eom} can be easily solved numerically. The results of the previous Section are essentially unchanged for flows initiated in the symmetric regime. For instance, in the massless, minimally coupled case, $m^2=\zeta=0$, one easily checks that Eqs.~\eqn{eqn:masslessIRinitial} and \eqn{eqn:masslessIR} for the initial and final values of $H_{\kappa}^2$ remain the same for all $N$, as illustrated in \Fig{fig:massless}. 

\begin{figure}[t]
    \centering
    \includegraphics[width=.4\textwidth]{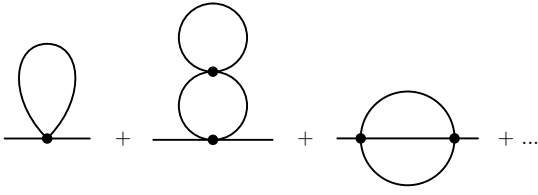}
    \caption{One- and two-loop diagrams contributing to the inverse correlator (self-energy) $\Omega\bar M_\kappa^2/H_\kappa^4$ (up to an overall sign) in the zero-dimensional theory \eqn{eqn:flowsol} in the symmetric regime. The lines represent the Gaussian correlator $\ev{\hat\varphi^a\hat\varphi^b}_{0,\kappa}=2\delta^{ab}\ev{\hat\rho}_{0,\kappa}$ and the vertices (dots) are given by $-\lambda \Omega/(8NH_\kappa^4)$.}
    \label{fig:self}
\end{figure}

\begin{figure}[t]
    \centering
    \includegraphics[width=.48\textwidth]{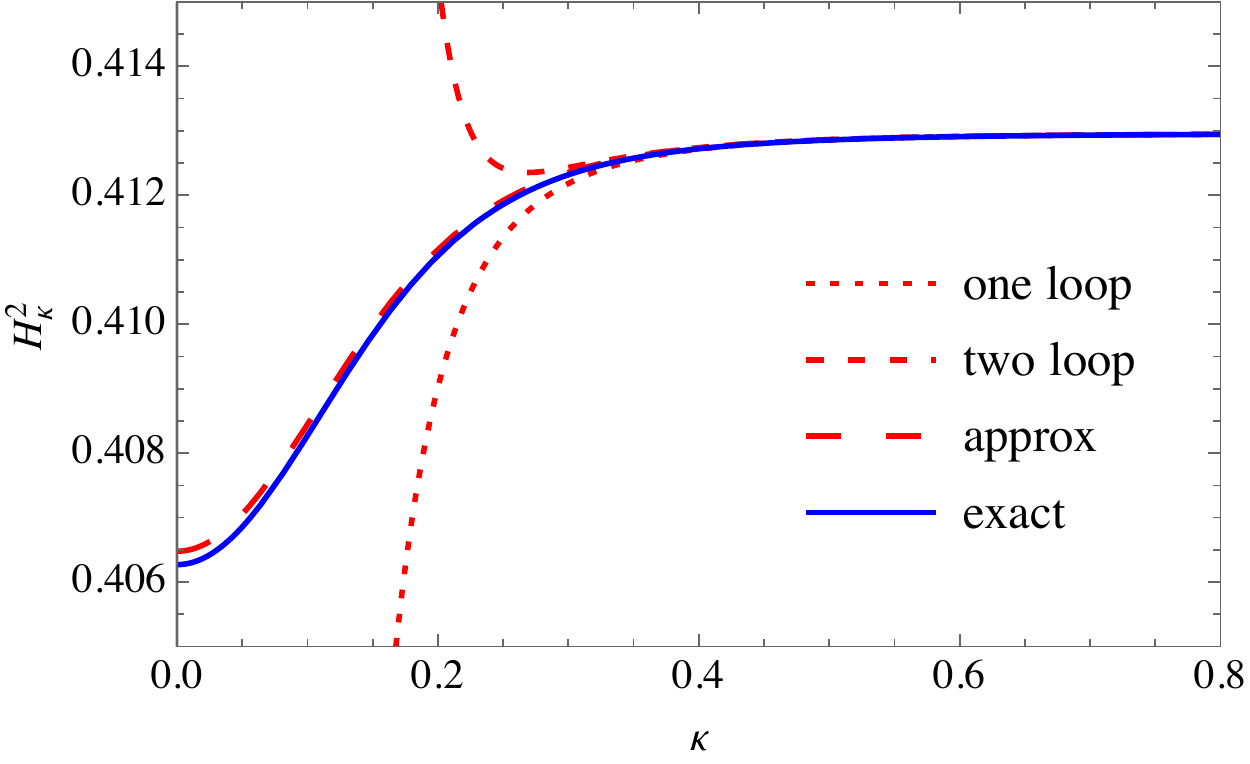}\\
    \vspace{.2cm}
    \hspace{.2cm}\includegraphics[width=.47\textwidth]{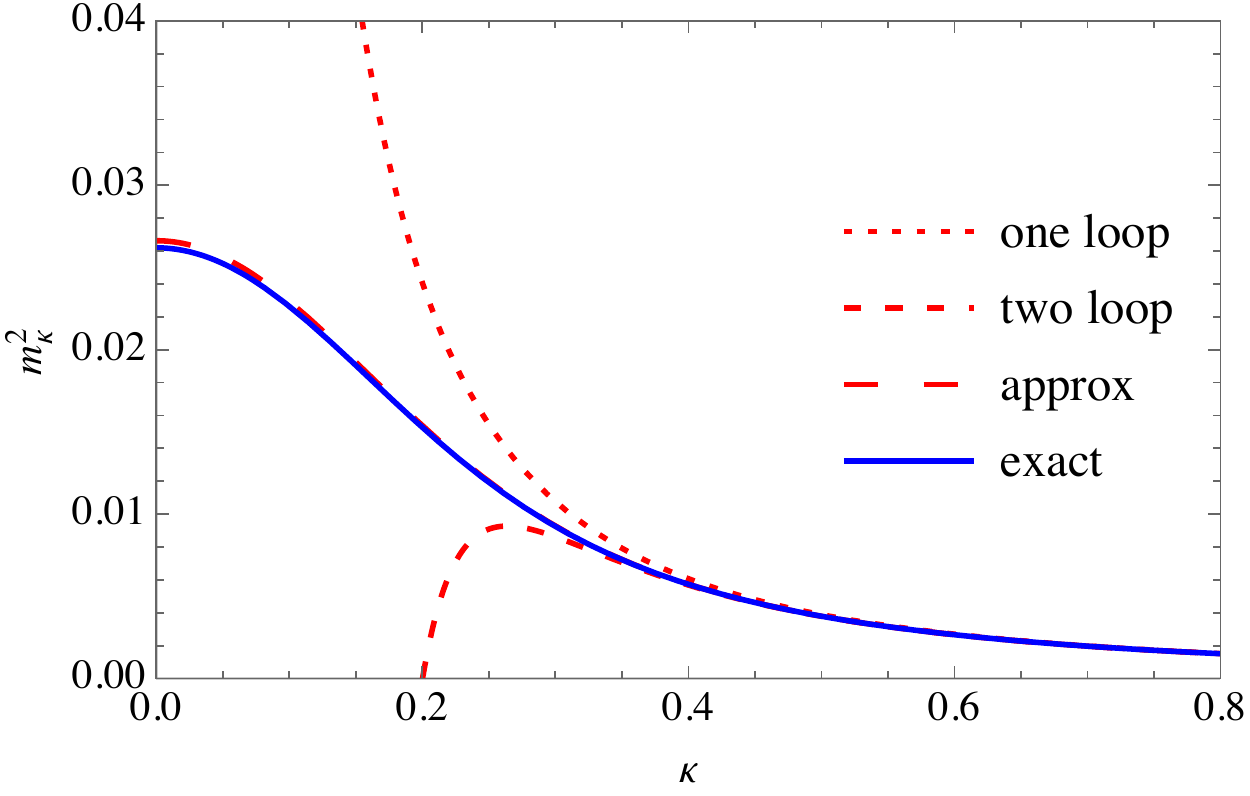}
    \caption{Flow of $H_\kappa^2$ (top) and of the running square mass $m_\kappa^2=\bar M_\kappa^2-\kappa^2$ (bottom) for the $N=1$ theory with $m^2=\zeta=0$ and $\lambda=0.1$. The gravitational parameters are $\alpha=0.1$ and $\beta=1$. Also shown are the one- and two-loop perturbative contributions, which correctly describe the flow at sufficiently large $\kappa$. For $\kappa^2\sim\sqrt{\lambda/\Omega}H_{\kappa_0}^2$ perturbation theory breaks down since all orders contribute equally. The flow of $H_\kappa$ eventually freezes as a nonperturbative mass is dynamically generated. The long-dashed curves show $\bar M^2_{\kappa,H_{\kappa_0}}-\kappa^2$ (bottom) and the approximate expression \eqn{eq:pertresum} (top).}
    \label{fig:loops}
\end{figure}

It is instructive to analyse the flow by means of perturbation theory \cite{Moreau:2018lmz}. In particular, for $m^2=\zeta=0$, where the tree-level correlator $\ev{\hat\rho}_{0,\kappa}=H_\kappa^4/(2\Omega\kappa^2)$, one would expect loop contributions to grow unbounded as $\kappa$ is decreased. For large enough $\kappa$, though, perturbation theory makes sense and one can use, {\it e.g.}, standard Feynman diagrams of the zero-dimensional theory \eqn{eqn:flowsol}, as represented in \Fig{fig:self}. It is easy to convince oneself that the actual expansion parameter is $\lambda \Omega \!\ev{\hat\rho}_{0,\kappa}^2/H_\kappa^4=\lambda H_\kappa^4/(4\Omega\kappa^4)=\lambda_{{\rm eff},\kappa}/4$, which grows with decreasing $\kappa$ as a direct result of the amplification of the tree-level correlator. In the symmetric regime, the one-loop order contribution to the the self-energy $\bar M^2_{\kappa}=\bar M^2_{t,\kappa}=\bar M^2_{l,\kappa}$ is given by the first diagram of \Fig{fig:self}, which yields
\beq
 \bar M_\kappa^2=\kappa^2+\frac{N+2}{2N}\frac{\lambda H_\kappa^4}{\Omega\kappa^2}+{\cal O}\qty(\lambda_{{\rm eff},\kappa}^2)
\eeq
or, equivalently, for the correlator,
\beq
 \ev{\hat \rho}_\kappa=\frac{H_\kappa^4}{2\Omega\kappa^2}\qty(1-\frac{N+2}{2N}\frac{\lambda H_\kappa^4}{\Omega\kappa^4}+{\cal O}\qty(\lambda_{{\rm eff},\kappa}^2))
\eeq
The equation for $H_\kappa$,
\beq\label{eq:Hmassless}
    4\alpha-\beta H_\kappa^2 + \frac{H_\kappa^4}{\Omega}\qty(1+\frac{\kappa^2}{\bar M^2_\kappa}) = 0 ,
\eeq
is solved by 
\begin{align}
 \frac{H_\kappa^2}{H_{\kappa_0}^2}&= 1-\frac{N+2}{2N}\frac{H_{\kappa_0}^2}{\beta\Omega-4H_{\kappa_0}^2}\frac{\lambda H_{\kappa_0}^4}{\Omega\kappa^4}+{\cal O}\qty(\lambda_{{\rm eff},\kappa}^2)\nn
 \label{eq:perturbative}
 &\approx 1-\frac{N+2}{2N}\frac{H_{\kappa_0}^2}{\beta\Omega}\frac{\lambda H_{\kappa_0}^4}{\Omega\kappa^4}+{\cal O}\qty(\lambda_{{\rm eff},\kappa}^2),
\end{align}
where we have used $H_{\kappa_0}^2\ll \beta$ in the second line. We thus see that the infrared one-loop contribution decreases the effective spacetime curvature. This could be interpreted as a sign of a possible instability of de Sitter space against loop corrections since, as it stands, the perturbative result \eqn{eq:perturbative} tends to rapidly drive $H_\kappa$ to zero as one integrates more and more amplified infrared fluctuations. However, for values of $\kappa$ where the one-loop correction becomes significant, the perturbative expansion parameter is not small anymore and all loop contribute equally significantly, as illustrated in \Fig{fig:loops}. As we have seen in the previous Section, nonpertubative effects actually generate a dynamical mass which screens the growth of infrared fluctuations ans freezes the flow of $H_\kappa$. 

The nonperturbative running square mass $\bar M_\kappa^2$ can be expressed in terms of special functions using the integral representation \eqn{eqn:flowsol}. For instance, in the symmetric regime, defining
\beq
 Z(A,B)=\int_0^\infty d\hat\rho \,\hat\rho^{N/2-1}e^{-A\hat\rho-\frac{B\hat\rho^2}{2}},
\eeq
one has
\beq
 \ev{\hat\rho}_\kappa =-\partial_A\ln Z(A,B),
\eeq
where the right-hand side must be evaluated at $A=N{\cal V}_D\bar\mu_\kappa^2$ and $B=N{\cal V}_D\lambda$. We get
\beq\label{eq:rhoexact}
 \frac{\ev{\hat\rho}_\kappa}{\ev{\hat\rho}_{0,\kappa}}=\frac{\bar\mu_\kappa^2}{\bar M_\kappa^2}=\frac{N}{2\lambda^{\rm eff}_\kappa}\frac{U\qty(\frac{N+4}{4},{3\over2},\frac{N}{2\lambda^{\rm eff}_\kappa})}{U\qty(\frac{N}{4},{1\over2},\frac{N}{2\lambda^{\rm eff}_\kappa})}
\eeq
where $\lambda^{\rm eff}_\kappa=\lambda H_\kappa^D/(\Omega\bar\mu_\kappa^4)$ has been introduced before and where $U(a,b,c)$ is the confluent hypergeometric function of the second kind. This expressions simplifies both in the limit $N\to\infty$, discussed above, and in the case $N=1$, discussed in \cite{Moreau:2018lmz}. One recovers the known expression for the generated mass of the massless, minimally coupled theory ($m^2=\zeta=0$) \cite{Starobinsky:1994bd,Guilleux:2015pma}:
\beq
 \frac{\bar M_{\kappa=0}^2}{H_{\kappa=0}^2}=\frac{\Gamma\qty({N\over4})}{\Gamma\qty({N+2\over4})}\sqrt{\frac{\lambda N}{8\Omega}},
\eeq
where the nonanalytic $\lambda$ dependence signs the intrinsic nonperturbative nature of the phenomenon.

As already mentioned, \Eqn{eq:Hmassless} is an implicit equation for $H_\kappa$ due to the nontrivial $H_\kappa$ dependence in \Eqn{eq:rhoexact}. However, we can obtain an explicit approximate solution by expanding \Eqn{eq:Hmassless} in inverse powers of $\tilde\beta=\beta\Omega/H_{\kappa_0}^2$ around $H_{\kappa_0}$, taking advantage of the small renormalisation effects. We obtain, for $m^2=\zeta=0$,
\beq\label{eq:pertresum}
 \frac{H_\kappa^2}{H_{\kappa_0}^2}=1-\frac{H_{\kappa_0}^2}{\beta\Omega}\qty(1-\frac{\kappa^2}{\bar M^2_{\kappa,H_{\kappa_0}}})+{\cal O}(\tilde\beta^{-2}),
\eeq
where the running mass on the right-hand side is given by \Eqn{eq:rhoexact} evaluated at $H_\kappa\to H_{\kappa_0}$. Note that, for large enough values of $\kappa$, where the perturbative treatment is valid, \Eqn{eq:pertresum} relates the loop expansion of $H_\kappa^2$ to that of $\bar M_\kappa^2$. However, \Eqn{eq:pertresum} remains valid in the nonperturbative regime. This is illustrated in \Fig{fig:loops}, together with the breakdown of the perturbative expansion, for $N=1$. 

Finally, for initial conditions in the broken symmetry regime, the main qualitative change as compared to the case $N=\infty$ studied above is that the longitudinal mode gives a nontrivial contribution to the flow of $H_\kappa^2$, as shown in \Fig{fig:rolegoldstone}. In this regime, the effect of the coupling $\lambda$ only appears at two-loop order and yields a positive renormalisation. Again, the symmetry gets restored at a finite RG scale $\kappa_*$ below which the flow is that of the symmetric regime, where the field selfinteraction drives a negative renormalisation of $H_\kappa$. We also observe, as before, that a negative nonminimal coupling $\zeta$ has an opposite effect as that of $\lambda$ and can change the sign of the renormalisation of $H_\kappa$ both in the broken symmetry and in the symmetric regime. 
\begin{figure}[t]
    \centering
    \includegraphics[width=.48\textwidth]{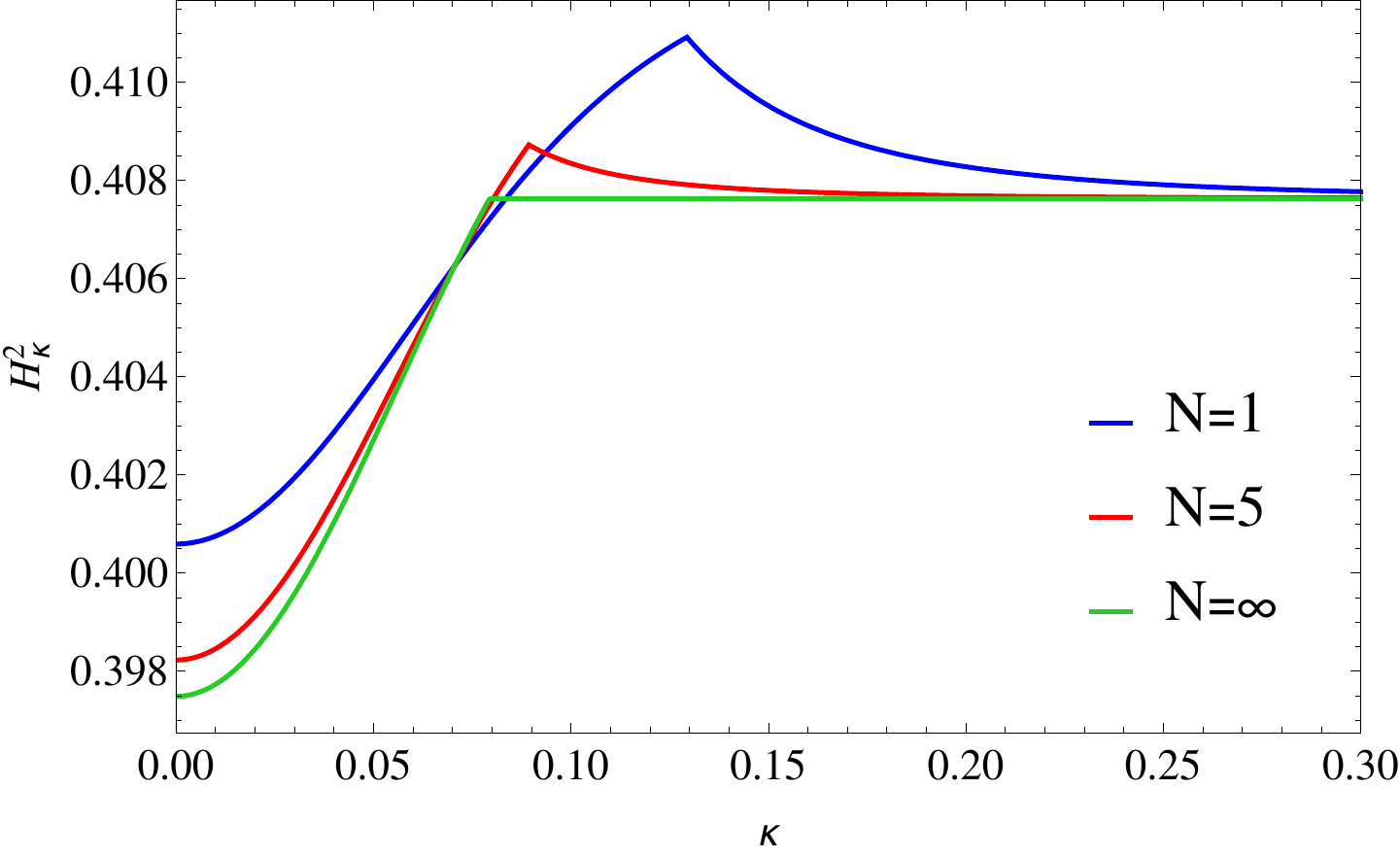}
    \caption{Flow of $H_\kappa^2$ for various values of $N$, with $m^2=-5\times10^{-3}$, $\zeta=0$, and $\lambda=10^{-2}$. The role of the massive longitudinal mode in the broken symmetry regime is clearly visible. The cusp corresponds to the scale at which the symmetry gets restored by the strong infrared fluctuations. The gravitational parameters are $\alpha=0.1$ and $\beta=1$.}
    \label{fig:rolegoldstone}
\end{figure}

\section{Conclusions}
\label{sec:concl}

To summarise, we have set up a NPRG formulation of semiclassical gravity, which we have used to study the backreaction of quantum scalar fields on a classical de Sitter geometry. We progressively integrate the nonperturbative superhorizon fluctuations of light fields and we study the resulting effective renormalisation of the spacetime curvature $\propto H_\kappa^2$. We verify that the theory of a noninteracting, minimally coupled field is a fixed point of the RG flow, whereas either a nonminimal coupling $\zeta$ to the Ricci scalar or a nonzero self-interaction $\lambda$ trigger a nontrivial flow. The former can lead to a positive or negative renormalisation of $H_\kappa$ depending  on the sign of the coupling, while the latter typically tends to decrease $H_\kappa$ in the infrared (in the symmetric regime). One striking result of the present study is that massless, minimally coupled fields (corresponding to exactly flat directions in the effective potential), despite being strongly amplified by gravitational effects, do not contribute to the infrared flow of $H_\kappa$. This is, in particular, the case of Goldstone modes in the broken symmetry regime. In that case the flow is controlled by the longitudinal mode and is thus suppressed for larger values of $N$. In all cases, spontaneously  broken symmetries at the initial scale $\kappa_0$ get restored by gravitationally amplified fluctuations in the infrared and the remaining flow is that of the symmetric regime. 

For the paradigmatic case of massless, minimally coupled fields, the large infrared quantum fluctuations triggered by the gravitational field lead to growing loop corrections which tend to rapidly drive $H_\kappa$ towards zero as $\kappa$ is decreased. However, when such loop contributions become important, the perturbative expansion breaks down and the flow is governed by nonperturbative effects. In particular, a nonzero mass is dynamically generated, which screens the large infrared fluctuations and leads to a saturation of the flow of $H_\kappa$ to a finite value. 

Our main findings are that, for generic parameters, the renormalisation of $H_\kappa$ due to the nonperturbatively amplified infrared fluctuations is controlled by the gravitational coupling $H_{\rm cl}^2/\beta\propto H_{\rm cl}^2/M_P^2\ll1$, where $H_{\rm cl}$ is the classical Hubble parameter either in the symmetric or broken symmetry case ({\it i.e.}, $H_{\rm cl}'$ in that case). Although it appears possible to fine-tune the parameters so that the infrared renormalisation of $H_\kappa$ be significant, we find no case where the spacetime curvature is renormalised to zero. In all cases, the dynamically generated square mass is $\bar M_{\kappa=0}^2\ll H_{\kappa=0}^2$, which ensures that the present flow equations remain valid all the way to $\kappa=0$.

The present work can be extended in various directions. It would be interesting to study the onset of gravitational effects as the flow of $H_\kappa$ is integrated from subhorizon scales, where the physics is that of Minkowski spacetime, to the regime of superhorizon scales studied here. This may require approximations beyond the LPA as derivative terms in the energy-momentum tensor as the latter do contribute to the subhorizon part of the flow. Another possible extension is to apply the present approach to less symmetric spacetimes, for instance, including slow-roll corrections to de Sitter spacetime. Finally, it is interesting to investigate the possibility of applying the present NPRG approach to similar backreaction problems, such as, the Schwinger effect in a constant electric field, or the black-hole evaporation due to Unruh-Hawking radiation.

Finally, although we believe that our work brings an interesting light on the question of the stability of de Sitter spacetime against quantum fluctuations, it remains limited in scope in that we restrict to the effect of scalar field fluctuations in a truly de Sitter invariant state. This clearly does not address a lot of important questions raised in the literature, such as the question of stability of a global de Sitter geometry (as opposed to the expanding Poincar\'e patch studied here) \cite{Polyakov:2007mm,Akhmedov:2012dn}, the role of non-de Sitter-symmetric quantum states for the scalar field \cite{Boyanovsky:2011xn,Anderson:2013ila,Akhmedov:2013xka,Anderson:2017hts,Akhmedov:2017ooy}, or the issue of graviton flucutations \cite{Tsamis:1992sx,Miao:2017vly}. It remains to be studied whether these could be addressed within the NPRG framework presented here.

\end{document}